\documentclass[12pt]{article}
\usepackage{amssymb,latexsym}
\oddsidemargin=\evensidemargin
\advance\topmargin by-\headheight
\def\ker{\mathop{\rm ker}\nolimits}

\def\mapright#1{\smash{\mathop{\rightarrow}\limits^{#1}}}
\def\mod{\mathop{\rm mod}\nolimits}

\def\ZZ{{\mathbb Z}}

\newtheorem{formula}{}[section]

\newtheorem{definition}[formula]{Definition}
\newtheorem{corollary}[formula]{Corollary}
\newtheorem{remark}[formula]{Remark}
\newtheorem{lemma}[formula]{Lemma}
\newtheorem{theorem}[formula]{Theorem}

\def\thrm{\begin{theorem}}
\def\thrml#1{\begin{theorem}\label{#1}}
\def\ethrm{\end{theorem}}
\def\rmrk{\begin{remark}}
\def\rmrkl#1{\begin{remark}\label{#1}}
\def\ermrk{\end{remark}}
\def\dfntn{\begin{definition}}
\def\dfntnl#1{\begin{definition}\label{#1}}
\def\edfntn{\end{definition}}
\def\nmrt{\begin{enumerate}}
\def\enmrt{\end{enumerate}}

\def\qtn{\begin{equation}}
\def\qtnl#1{\begin{equation}\label{#1}}
\def\eqtn{\end{equation}}
\def\lmm{\begin{lemma}}
\def\lmml#1{\begin{lemma}\label{#1}}
\def\elmm{\end{lemma}}
\def\crllr{\begin{corollary}}
\def\crllrl#1{\begin{corollary}\label{#1}}
\def\ecrllr{\end{corollary}}

\begin{document}
\title{Public-key cryptography and invariant theory}
\author{
Dima Grigoriev \\[-1pt]
\small IRMAR, Universit\'e de Rennes \\[-3pt]
\small Beaulieu, 35042, Rennes, France\\[-3pt]
{\tt \small dima@maths.univ-rennes1.fr }
}
\date{}
\maketitle

\begin{abstract}
Public-key cryptosystems are suggested based on invariants of groups. We give
also an overview of known cryptosystems which involve groups.

\end{abstract}

\section{Introduction}

In public-key cryptography the problem is to produce a cryptosystem which
contains the following ingredients: a public key $k_e$, a secret key $k_d$,
a public encrypting function $f_e$ and a secret decrypting function $f_d$.
If somebody (usually named Bob) wants to send a message $m$ to another person
(usually named Alice) via a public channel then he transmits an encryption
$u=f_e(m,k_e)$. To decrypt a message Alice calculates $m=f_d(u,k_d)$. It is
supposed that $k_d, f_d$ are known only to Alice, while $k_e, f_e$ are known
publically. Another important property of a cryptosystem is that an unauthorized person (named Charlie) would be unable to learn $m$ from $u$ (without knowing
$k_d, f_d$).

A lot of efforts were undertaken to design cryptosystems (some literature one
can find in \cite{K,GB,G}). Still for no cryptosystem its
security is proved and the issue of security remains a challenging problem.
All the existing results on security concern impossibility of breaking a
cryptosystem by certain fixed means, say, in frames of a particular proof system.
But what appeared to be interesting in the course of developpment of cryptography is that many connections with other areas of mathematics were discovered.
In fact, public-key cryptography in many aspects plays a role of a bridge
between mathematics and computer science.
The
most recognized cryptosystems are based on number-theoretical ideas like RSA,
Diffie-Hellman or the elliptic curves cryptosystems (see e.g., \cite{K}, in this book one can find also some cryptosystems invoking combinatorial-algebraic
NP-hard problems).

In these notes we study cryptosystems which involve ideas from the theory of
group invariants. Several known cryptosystems rely on groups (below we give a
short overview of them), but surprisingly the concept of a group representation invariant was never exploited, although it fits quite well the aims of
cryptography. Still, applying invariants into cryptography encounters the
similar difficulties as other known approaches, however, the purpose of these notes
is to introduce invariants into cryptography, design a possible cryptosystem
and to discuss arising problems on its security.

Thus, the general idea behind using groups (and their invariants) is as follows. Let $E$ denote the set of {\it encrypted messages} and a group $G$ act $G:E \rightarrow E$. Some examples of $E$ are the set of words in a certain alphabet or a
vector space. In addition, a subset $M \subset E$ of {\it plaintext messages} is
distinguished being transversal to the orbits of $G$, i.e. no two distinct
messages $m_1,m_2 \in M$ belong to the same orbit of $G$.

Then $G$ is used for a {\it probabilistic} encryption \cite{GM,GB} being more
efficient in ``hiding information''  than a deterministic one. Thus, for
encrypting a message $m\in M$ Bob picks randomly an element $g$ from the group
$G$ and transmits $gm \in E$ via the public channel. Alice has to decrypt $gm$
to learn $m$ applying her secret key. And Charlie has to be unable to learn
$m$ from $gm$.

Usually, the latter two properties of the cryptosystem are achieved by a
special choice of $G$. In the widely used quadratic residue cryptosystem
\cite{GM,GB} one takes an integer $n=pq$ for large primes $p,q$ (being a secret
key of Alice). As the group $G$ one takes $(\ZZ_n^*)^2$ and
$E=\{g\in \ZZ_n^*: J_n(g)=1\}$ where $J_n$ denotes the Jacobi symbol. Take also
an element $a\in E-G$ (being a non-square), put $M=\{1,a\}$. A public key
consists of $n$ and $a$. Thus, to encrypt 1 Bob picks first a random $g\in \ZZ_n^*$, then its square $g^2 \in G$, and to encrypt $a$ Bob picks $g^2a$ being a
random non-square in $E$, clearly $E=G \cup Ga$.

The task of Alice is to verify whether an element $b \in E$ (being a transmitted encrypted message) is a square. This Alice can easily do using $p,q$ and
 Legendre-Jacobi symbols $J_p, J_q$. On the other hand, it is a common
belief that Charlie is unable to verify whether $b$ is a square without knowing
$p,q$.

The described quadratic residue cryptosystem was generalized to a class of
cryptosystems called homomorphic. Namely, let $f:E \rightarrow H$ be a group
epimorphism which is a secret key of Alice. There is an exact sequence of
group homomorphisms

$$B\, \mapright{s}\, E\, \mapright{f}\, H\, \mapright{}\, 0$$

\noindent
and a public key is $B,s,E,H$ and a subset $M \subset E$ transversal to $G=\ker(f)$, hence $f$ provides a bijection between $M$ and $H$. This is consistent with
the above notations: $G$ acts (by the multiplication from the left) on $E$ and the set of plaintext messages $M$ being transversal to this action, but here
$G$ is given implicitly as an image $s(B)$. To encrypt a message $m \in M$ Bob picks randomly $b\in B$ and transmits $s(b)m$. Alice decrypts $s(b)m$ applying
$f$, taking into account that $f(s(b)m)=f(m)$. For Charlie it is difficult to
decrypt without knowing $f$. In the described above quadratic residue
cryptosystem we have $H=\ZZ_2, B=E, s(b)=b^2$ and $f$ being the epimorphism of
the quadratic residue.

In \cite{FM,Y} a question was posed for which groups $H$ homomorphic systems
can be constructed (more generally, one could consider rings rather than groups homomorphisms)? For some abelian groups $H$ cryptosystems were designed in
\cite{B,NS97,NS,OU98,P99}. For certain diedral groups $H$ cryptosystems are
designed in \cite{R00}. In \cite{GP} a homomorphic system was designed for
any solvable group $H$. For cryptosystems over elliptic curves see \cite{KMOV,K}.

What is common in all the mentioned constructions is that decrypting relies on
the knowledge of secret primes $p,q$. In these notes we suggest another way of
decrypting (and encrypting) based on an invariant $w:E\rightarrow F$, i.e. $w$ being constant
on the orbits of $G$. Then Alice is able to decrypt an encrypted message $gm$
by means of calculating $w(gm)=w(m)$, provided that $w$ takes distinct values
on the elements (plaintext messages) from $M$.

The theory of invariants, see e.g. \cite{DC,S77,S93}, is developped mostly in the situation when $G:E\rightarrow E$ is a linear representation, so $E$ is a vector space and
$G \subset GL(E)$ over a field $F$ and $w$ being a polynomial. But perhaps, it
would be also worthwhile to look at other group actions and their invariants.

Since for not too many infinite series of linear representations $G \subset
GL(E)$ their invariants $w$ are known explicitly and can be calculated fast (\cite{S93}),
we suggest to hide $G$ considering its conjugation $a^{-1}Ga \subset GL(E)$ for a secret matrix $a\in GL(E)$. Then an invariant $e\rightarrow w(ae)$ of the
conjugation $a^{-1}Ga$ enables Alice to decrypt a message $a^{-1}gam$ where
$a^{-1}ga$ is a random element from $a^{-1}Ga$ chosen by Bob for encrypting a
message $m$. Usually, the group $a^{-1}Ga$ (being a public key together with
$E$ and $M\subset E$) is given by a set of matrices in $GL(E)$ being its
generators. It is a quite succinct way of representing a group by a set of its
generators, in particular, known finite simple groups are representable just by two
generators, and any finite group $G$ is representable by $\log _2|G|$ generators. In calculations with $G$ represented by a set of generators it is not
necessary to assume that $G$ is finite (which is the case in particular, when the field $F$ is finite) because for encrypting Bob has just to pick randomly a
certain product of generators of $a^{-1}Ga$.

In the next sections we describe cryptosystems
based on group invariants and discuss the issues of their security, but first we
complete an overview by two families of cryptographic tools which involve groups.

Another particular problem of cryptography, apart from designing cryptosystems and
closely connected with it, is the {\it key agreement protocol}, see e.g. \cite{GB,G,K}. Now Alice and Bob want to agree about a common key communicating via a public channel. The usual approach is to choose by each of them secretly
commutating operators $f_A$ (by Alice) and $f_B$ (by Bob) in the same set $E$
and in addition a certain (public) $e\in E$. Then Alice communicates $f_A(e)$,
Bob communicates $f_B(e)$ and they agree on a common key $f_A(f_B(e))=f_B(f_A(e))$. In the first key agreement protocol due to Diffie-Hellman (see e.g. \cite{GB,K}) it was used $f_A(e)=e^a, f_B(e)=e^b (\mod p)$ for integers $a,b$. Thus, decrypting (by Charlie) of Diffie-Hellman protocol relates to computing the discrete
logarithm which is believed to be difficult, its complexity was studied in
\cite{Sp,MW}.

This general approach was considered in the following setting (see \cite{AAG,PKHK,KL}. Let $E$ be a group with two subgroups $E_A,E_B \subset E$ which commute with each other. Then as $f_A$ Alice chooses a conjugation $e\rightarrow a^{-1}ea$
for a randomly picked $a\in E_A$, respectively, $f_B(e)=b^{-1}eb$ for $b\in E_B$. In \cite{KL} the braid group is used as $E$ and the difficulty of breaking
this key agreement protocol relates to the difficulty of the conjugacy problem
in the braid group.

Few cryptosystems based on the difficulty of the word problem in appropriate
groups were proposed in \cite{AA,DJSS,GZ,WM}.

The last family of cryptosystems we mention rely on lattices (being discrete
abelian subgroups of ${\mathbb R}^n$), the first such a construction is due to
\cite{AD}. Let $L \subset {\mathbb R}^n$ be an $n$-dimensional lattice with a
property that it contains a (hidden) $(n-1)$-dimensional sublattice $L'$ whose
linear span being a hyperplane $H$, satisfying the following property. The
coset hyperplanes $H_i$ parallel to $H$ such that $\cup_i H_i \supset L$ are
well separated: the distance between any adjacent pair $H_i,H_{i+1}$ of them is
greater than a suitable large $d$. In other terms, there is a basis of $L$
which consists of $n-1$ rather ``short'' vectors which form a set $C$ from $L'$
and a single ``long'' vector $c$ having a ``big'' coordinate orthogonal to $H$.

Then a cryptosystem from \cite{AD} considers a random basis $C_1$ of $L$ as a
public key and a basis $C \cup \{c\}$ as a secret key. The plaintext 0 is
encrypted by Bob by a random vector from $L$, and the plaintext 1 is encrypted by a
random vector from  ${\mathbb R}^n$.

For decrypting a vector $u$ Alice computes a magnitude $l=(u,c-P_H(c))/||c-P_H(c)||^2$ where $P_H$ denotes the orthogonal projector onto the hyperplane $H$. If $l$ is an integer (this means that $u$ lies on a certain coset hyperplane
$H_i$) then Alice can declare that the plaintext message of Bob is 0 (otherwise,
1). Actually, in this manner Alice recognizes elements from $\cup_i H_i$ rather
than from $L$, so an error happens when $u \in \cup_i H_i -L$. To correct this
Bob slightly perturbs $u$, so each point from $L$ is surrounded by a ball of
a suitable radius $r$ in order to cover $\cup_i H_i$, but on the other hand, not
to cover the whole space ${\mathbb R}^n$. Moreover, the perturbations of adjacent
hyperplanes $H_i$ and $H_{i+1}$ should be well separated, just for this reason
the condition on a ``long'' vector $c$ was imposed. Finally, Alice decrypts the
points at the distance at most $r$ from the union $\cup_i H_i$ as 0, otherwise as 1. Still, an error could
happen when $u$ lies at the distance at most $r$ from the union $\cup_i H_i$, but
now it is more probable that Bob encrypted 0 in this case (rather than 1).

Thus, the presumed difficulty of breaking (by Charlie) this cryptosystem relies on finding a
long vector in a lattice given by its basis $C_1$ (or equivalently, a short one in the dual lattice),
provided that a long vector is unique in an appropriate strong sense.

Another cryptosystem based on perturbations of a lattice was designed in \cite{GGH}. Here a plaintext message is a point of a lattice $L\subset {\mathbb R}^n$ and
its encryption is its small perturbation in ${\mathbb R}^n$. Then the problem
of breaking the cryptosystem leads to finding for a given real point the closest to it vector in the lattice $L$. This problem is known to be NP-hard, as well as
its approximating up to a constant factor. To make the decrypting possible Alice
first chooses (randomly) a basis $c_1,\dots , c_n$ of $L$ of a special form
(namely, such that the magnitude $\prod_i ||c_i||/|\det(c_i)|$ is not too large,
a basis with this property is called ``almost rectangular'') which serves as a
secret key. After that the basis $c_1,\dots , c_n$ is (randomly) spoiled and the
resulting new basis of $L$ serves as a public key. The point is that an almost
rectangular basis allows Alice to find the closest vector in $L$, provided that
a perturbation was small enough.

Thus, in both mentioned lattice-based methods \cite{AD,GGH} a plaintext message
is hidden by a small perturbation, this differs from our suggestion to hide
by means of shifting by an element from a certain group.

\section{Construction of cryptosystems based on group invariants}

Let $G\subset GL_n(F)$ be a representation of a group $G$ where one can deem
w.l.o.g. a field $F$ to be algebraically closed, however in computations the
entries of the matrices from $G$ could belong to a certain subfield of $F$, it is reasonable, for example, the entries to belong to a finite subfield.

Assume that we know a (non-constant) invariant $w$ (\cite{DC,S77}) of the
representation of $G$, i.e. a polynomial $w\in F[X_1,\dots , X_n]$ such that
for any element $g\in G$ and any vector $v\in F^n$ we have $w(gv)=w(v)$. Besides, we fix a pair of nonzero distinct vectors $v_0,v_1 \in F^n$.

Usually (and we suppose this) one is able to generate elements from $G$. To
design a (probabilistic) public-key cryptosystem Alice chooses randomly a
matrix $a\in GL_n(F)$ with the property that $w(av_0)\neq w(av_1)$ (clearly,
almost any matrix $a$ satisfies this property).

\vspace{4mm}
\noindent
{\bf Public key}: $v_0,v_1$ and a set of elements of the form $h_i=a^{-1}g_ia \in GL_n(F)$ where $g_i$ being randomly generated elements of $G$.

\vspace{2mm}
\noindent
{\bf Secret key}: $a$

\vspace{2mm}
\noindent
{\bf Encryption}: a letter 0 or respectively, 1 of a plaintext message is transmitted as a
vector $u=h_{i_1}\cdots h_{i_l}v_0$ (or respectively, $u=h_{i_1}\cdots h_{i_l}v_1$) for randomly chosen $i_1,\dots , i_l$.

\vspace{2mm}
\noindent
{\bf Decryption}: given a vector $u\in F^n$ Alice computes $w(au)$ and verifies
whether it equals to $w(av_0)$ (in this case the plaintext message was $v_0$
since $w(av_0)=w(ah_{i_1}\cdots h_{i_l}v_0)=w(au)$) or to $w(av_1)$ (in this
case the plaintext message was $v_1$).

\section{Discussion on the security}

To break the designed cryptosystem Charlie can try to find a certain invariant
$w'$ of a (sub)group $H$ of a conjugation $a^{-1}Ga$ (where $H$ is given by a
set of generators $h_i$). One can think of the group $G\subset GL_n(F)$ (of
exponential in $n$ size or even infinite) to be known as well as an invariant $w$. Charlie could try to look for an invariant $w'$ in the form $w'(v)=w(bv)$ for
an unknown matrix $b\in GL_n(F)$. Let $d$ denote the degree of $w$ (as usually,
in the invariant theory one might reduce consideration to a homogeneous $w$).
Substituting $w'$ into the known generators of the form $a^{-1}g_ia$ of $H$, so
$w(bv)=w(b(a^{-1}g_ia)v)$ for each $i$ (clearly, any such matrix $b$ would fit),
and equating the coefficients at all the monomials in $n$ coordinates of a
vector $v$, Charlie obtains a system of polynomial equations in the entries of a
matrix $b$ of degrees $d$. Hence this polynomial system contains ${n+d-1}\choose
{d}$ equations (for each $i$) of degrees $d$ in $n^2$ variables being the
entries of $b$.

Alternatively, Charlie could search for an invariant $w'$ treating it as a
polynomial of degree $d$ with indeterminate ${n+d-1}\choose
{d}$ coefficients satisfying equations $w'(h_iv)=w'(v)$ for each $i$ for any
vector $v\in F^n$. This provides a linear system in the indeterminate
coefficients.

Anyway, the complexity of both procedures depends on  ${n+d-1}\choose
{d}$, therefore, for the security reasons one should take a group $G$ without
invariants of degrees $d$ less than $\mbox{const} \cdot n$. On the other hand, the
invariant $w$ should be computable within polynomial in $n$ complexity (below
we give few such examples).

The problem of finding an invariant seems to be difficult and in general not
much is known beyond obvious applying the Reynolds averaging operator
$|G|^{-1}\sum_{g\in G} g$ (provided that $G$ is finite), cf. \cite{S93}.

Let us consider two other approaches towards breaking the described cryptosystem.

In the first approach Charlie tries to find a matrix $a$ (or any other $b\in
GL_n(F)$ such that $bHb^{-1} \subset G$). Clearly, one can assume w.l.o.g. that
$H$ is conjugate to $G$ itself (rather than to a certain its subgroup) taking
as $g_i$ a set of generators of $G$ (randomly chosen large enough set of $g_i$
generates $G$ with a high probability). Then testing an existence of $b$ (and
finding if it does exist) such that $bHb^{-1}=G$ is called the {\it conjugacy
problem for matrix groups}. One can reduce to the latter problem {\it the conjugacy problem for permutation groups} as it was communicated to the author by Eugene Luks
\cite{L02}. In its turn, the difficulty of the conjugacy problem for
permutation groups was conjectured in \cite{L93} where its complexity was
posed as an open question.
Furthermore, the {\it graph isomorphism problem} is reducible to the
conjugacy problem for permutation groups \cite{L93}. Thus, the first approach by Charlie
leads in particular, to the graph isomorphism problem.

In the second approach Charlie tries to find a matrix $h$ from $H$ such that
$hu=v_0$ (or respectively, $hu=v_1$). This problem (in a particular case when
$H$ is a permutation group) is called the {\it vector transporter problem}
\cite{L93} where its difficulty was conjectured. Again the graph isomorphism
problem is reducible to the vector transporter problem \cite{L93}.

We observe that a particular case of the vector transporter problem when (the
set of encrypted messages) $F^4$ is the space of $2 \times 2$ matrices and
a group $H=SL_2(\ZZ) \times SL_2(\ZZ)$ acts on $F^4$ by $v\rightarrow h_1vh_2$,
where $(h_1,h_2)\in H$, was proved to be NP-hard for the {\it average complexity} \cite{BG}. In this connection we mention that a hardness of breaking a
cryptosystem for the average complexity would be more desirable than a
hardness for the worst-case complexity (cf. \cite{L86}).

Thus, both approaches towards breaking the described above cryptosystem relate
to the graph isomorphism problem. But of course, the difficulty of proving
a reduction of the graph isomorphism to breaking the cryptosystem lies
particularly, in finding an appropriate invariant $w$ for a group $G$ such that
the graph isomorphism problem can be reduced to the conjugacy and to the
vector transporter problems for $G$. It would be interesting to understand
whether the graph isomorphism problem indeed, is reducible to breaking the
described above cryptosystem.

We mention also that a similarly looking problem of equivalence of
representations of the group algebras $F[G],F[H]$ (rather than the groups),
in other words finding a matrix $a\in GL_n(F)$ such that $a^{-1}F[H]a=F[G]$,
can be solved over an algebraically closed field $F$ \cite{R} taking into
account the structural theorems of Schur and Wedderburn.

An evident remark is that to provide more security of the cryptosystem
it would be reasonable to change secret and public keys $a$ and $a^{-1}g_ia$ quite often.

Let us give few simple examples of cryptosystems based on invariants of
classical groups \cite{DC,S77}.

\vspace{4mm}
{\bf Example 1} (\cite{S77}). As a group $G$ we take a subgroup of $GL_n$
generated by the symmetric group $S_n$ permuting the standard basis $e_i, 1\leq i \leq n$ and all the matrices $t$ such that $te_i=c_ie_i$ where $c_i^m=1,
1\leq i \leq n, (c_1\cdots c_n)^l=1$ for some $l|m$. Then as $w$ one can take
the power sum $x_1^m+\dots +x_n^m$. We deliberately consider an extension of
the symmetric group to avoid invariants of small degrees (see the beginning
of this section).

\vspace{2mm}
{\bf Example 2}. Consider the representation of the group $G=SL_n(F)$ on the
symmetric product $S^2F^n$, in other words, on symmetric matrices (or
quadratic forms) by $v\rightarrow mvm^T$ where $m\in G$ and $T$ denotes the
transposing. Then as $w$ one takes $\det (v)$.

\vspace{2mm}
{\bf Example 3}. Now $G=GL_n(F)$ which acts on the direct sum $F^{2n^2}=
F^n\oplus \cdots \oplus F^n$ of $2n$ copies of $F^n$ by $m(p_1,\dots , p_{2n})=
(mp_1,\dots , mp_{2n})$. Consider two (disjoint) partitions $I_1 \cup J_1=
I_2 \cup J_2 = \{1,\dots , 2n \}$ into $n$-element subsets $|I_1|=|J_1|=|I_2|=
|J_2|=n$. As $\det_{I_1}$ we denote the determinant  of $n$ vectors $p_i$ for
$i\in I_1$. Then as $w$ we take the {\it rational} invariant

$$\frac{\det_{I_1}\det_{J_1}}{\det_{I_2}\det_{J_2}}.$$

\noindent
In the described above cryptosystem we considered polynomial invariants $w$,
but nothing changes when we deal with rational invariants $w\in F(X_1,\dots ,
X_n)$, except that, of course, in the cryptosystem Alice should pick vectors
$v_0, v_1 \in F^{2n^2}$ and a matrix $a\in GL_{2n^2}(F)$ in such a way that
$w(av_0), w(av_1)$  be defined.

\vspace{4mm}
One could produce more similar examples invoking direct, tensor, symmetric,
exterior products of group representations.

The present state of art of cryptography does not allow to prove security
of cryptosystems, the latter is usually a question of belief in a difficulty of a
relevant problem and a matter of experience (that is why it is not quite
unusual to have a paper on cryptography without theorems, including this one).
Just the opposite, one could expect a ``disappointing'' breaking of a particular cryptosystem. This is not excluded for any of the aforementioned
examples (and avoiding solving the graph isomorphism problem, see the
discussion above). On the other hand, such breaking would lead perhaps, to
interesting algorithms in group representations. Thus, one can treat the
examples (and  the general construction in all)  just as a suggestion to play
with cryptosystems based on the invariant theory.

\vspace{5mm}

{\bf Acknowledgements.}
The author would like to thank the Institut des Hautes Etudes Scientifiques
during the stay in which this paper was conceived and also Lenya Levin and
Gene Luks for interesting discussions.

\end{document}